\documentclass[fleqn,usenatbib,usedcolumn]{mnras}

\usepackage[T1]{fontenc}
\usepackage{ae,aecompl}

\usepackage{graphicx}	
\usepackage{amsmath}	
\usepackage{amssymb}	

\newcommand{\um}{$\mu$m}

\title[Infrared polarimetry of Mrk 231]{Infrared polarimetry of Mrk 231: Scattering off hot dust grains in the central core}

\author[Lopez-Rodriguez et al.]{
E. Lopez-Rodriguez,$^{1,2}$\thanks{E-mail: \href{mailto:enrique.lopezrodriguez@utexas.edu}{enrique.lopezrodriguez@utexas.edu}},
C. Packham$^{3,4}$,    
T. J. Jones$^{5}$,  
R.  Siebenmorgen$^{6}$,
\newauthor 
P.~F.~Roche$^{7}$,
N.~A.~Levenson$^{8}$,
A. Alonso-Herrero$^{9,3}$, 
E. Perlman$^{10}$,
\newauthor  
K. Ichikawa$^{4}$,
C. Ramos Almeida$^{11,12}$,
O. Gonz\'alez-Mart\'in$^{13}$,
R. Nikutta$^{14}$,
\newauthor
M. Martinez-Paredez$^{13}$,
D. Shenoy$^{5}$, 
M.~S. Gordon$^{5}$,
C.~M. Telesco$^{15}$ 
\\
	$^{1}$Department of Astronomy, University of Texas at Austin, 1 University Station C1400, Austin, TX 78712, USA \\
	$^{2}$McDonald Observatory, University of Texas at Austin, Austin, TX 78712, USA \\
	$^{3}$Department of Physics \& Astronomy, University of Texas at San Antonio, One UTSA Circle, San Antonio, TX 78249, USA \\
	$^{4}$National Astronomical Observatory of Japan, 2-21-1 Osawa, Mitaka, Tokyo 181-8588, Japan \\
	$^{5}$Minnesota Institute for Astrophysics, University of Minnesota, 116 Church Street SE, Minneapolis, MN 55455, USA \\
	$^{6}$European Southern Observatory, Karl-Schwarzschild-Str. 2, 85748 Garching b. M\"unchen, Germany \\
	$^{7}$Astrophysics, Department of Physics, University of Oxford, DWB, Keble Road, Oxford OX1 3RH, UK \\
	$^{8}$Gemini Observatory, Casilla 603, La Serena Chile \\
	$^{9}$Centro de Astrobiologia, CSIC-INTA. ESAC Campus, E-28692 Villanueva de la Ca\~nada, Madrid, Spain  \\
	$^{10}$Florida Institute of Technology, Melbourne, FL 32901, USA.  \\
	$^{11}$Instituto de Astrof\'isica de Canarias, Calle V\'ia L\'actea s/n, 38205, Tenerife, Spain \\
	$^{12}$Universidad de La Laguna, Departamento de Astrof\'isica, E-38206 La Laguna, Tenerife, Spain \\ 		   
	$^{13}$Centro de Radioastronom\'ia y Astrof\'isica (CRyA-UNAM), 3-72 (Xangari), 8701 Morelia, Mexico \\
	$^{14}$Instituto de Astrof\'isica, Facultad de F\'isica, Pontificia Universidad Cat\'olica de Chile, 306, Santiago 22, Chile \\
	$^{15}$Department of Astronomy, University of Florida, 211 Bryant Space Science Center, Gainesville, FL 32611-2055, USA \\
	}

\date{Accepted XXX. Received YYY; in original form ZZZ}

\pubyear{2016}

\begin{document}
\label{firstpage}
\pagerange{\pageref{firstpage}--\pageref{lastpage}}
\maketitle

\begin{abstract}

We present high-angular (0.17$-$0.35 arcsec) resolution imaging polarimetric observations of Mrk 231 in the 3.1 \um~filter using MMT-Pol on the 6.5-m MMT, and in the 8.7 \um, 10.3 \um, and 11.6 \um~filters using CanariCam on the 10.4-m Gran Telescopio CANARIAS. In combination with already published observations, we compile the 1$-$12 \um~total and polarized nuclear spectral energy distribution (SED). The total flux SED in the central 400 pc is explained as the combination of 1) a hot (731 $\pm$ 4 K) dusty structure, directly irradiated by the central engine, which is at 1.6 $\pm$ 0.1 pc away and attributed to be in the pc-scale polar region, 2) an optically-thick, smooth and disk-like dusty structure (`torus') with an inclination of 48 $\pm$ 23\degr~surrounding the central engine, and 3) an extinguished (A$_{\mbox{\scriptsize V}} =$ 36 $\pm$ 5 mag) starburst component. The polarized SED decreases from 0.77 $\pm$ 0.14 per cent at 1.2 \um~to 0.31 $\pm$ 0.15 per cent at 11.6 \um~and follows a power-law function, $\lambda^{\sim0.57}$. The polarization angle remains constant ($\sim$108\degr) in the 1$-$12 \um~wavelength range. The dominant polarization mechanism is explained as scattering off hot dust grains in the pc-scale polar regions. 

\end{abstract}





\section{Introduction}
\label{INTRO}

Mrk 231 is classified as a Seyfert 1 \citep{Veilleux:1999aa}, the nearest (z=0.042, D=175 Mpc, 1 arcsec = 800 pc, using H$_{\mbox{\tiny o}}$ = 75.0 km s$^{-1}$ kpc$^{-1}$, $\Omega_{\mbox{\tiny m}}$ = 0.27, $\Omega_{\mbox{\tiny v}}$ = 0.73) broad absorption line quasar \citep{Boksenberg:1977aa,Rudy:1985ab}, and it is the most luminous object within 300 Mpc in the \textit{Infrared Astronomical Satellite} (\textit{IRAS}) survey \citep{Soifer:2000aa}. Mrk 231 shows an unresolved core \citep{Matthews:1987aa,Keto:1992aa,Lai:1998aa,Soifer:2000aa,Quillen:2001aa,Lonsdale:2003aa,Low:2004aa} in the near-infrared (NIR) to mid-IR (MIR) wavelength range with low visual extinction, $A_{\mbox{\tiny V}} \sim 2$ mag \citep{Boksenberg:1977aa,Jones:1989ab,Lipari:1994aa} and $A_{\mbox{\tiny V}} \sim 7$ mag \citep{Veilleux:2013aa} towards the region that obscures the Balmer lines produced in the broad line region (BLR). A silicate absorption band at 10 \um~with $\tau_{\mbox{\tiny 9.7}} \sim$ 1.2 indicates  much higher extinction to the MIR emitting region of A$_{\mbox{\tiny V}} \sim$ 20 mag \citep[e.g.][]{Roche:1983aa}. The upper-limit size of the core is as small as 0.008 arcsec (6 pc) at 1.1 \um~using \textit{Hubble Space Telescope} (\textit{HST}) observations \citep{Low:2004aa}, and 0.13 arcsec (104 pc) at 12.5 \um~using the 10.0-m Keck Telescope \citep{Soifer:2000aa}. At the same location as the MIR core, a compact, $\le$90 pc $\sim$ 0.1 arcsec, vibrationally excited HCN emission at mm wavelengths was detected \citep{Aalto:2012aa} consistent with a hot, dusty, warped inner disk.

Many efforts have been made toward understanding the physical structures in the core of Mrk 231. Polarimetric techniques offer a powerful tool as they are able to provide information far below the angular resolution of the observations. A multi-wavelength study aimed to disentangle the wavelength dependence of each polarization mechanism is then crucial. The ultraviolet (UV) to NIR continuum polarization of the unresolved core decreases from as high as 20 per cent in the UV \citep{Smith:1995aa} to as low as 0.5 per cent in the NIR \citep{Jones:1989ab}. This result is interpreted as dust scattering with modest starlight dilution within the unresolved core. Further MIR polarization observations \citep{Siebenmorgen:2001ab} using ISOCAM found a highly polarized (8.6 $\pm$ 0.9 per cent at 12.0 \um~and 6.7 $\pm$ 0.9 per cent at 14.3 \um) nucleus in a 9.6 arcsec (7.7 kpc) aperture. These authors suggested that the dominant polarization mechanism arises from the passage of radiation from the central engine through magnetically aligned dust grains in the torus. The position angle (P.A.) of polarization of the core rotates from $\sim$130\degr~at far-UV to $\sim$90\degr~in the optical, remaining almost constant up to the NIR \citep{Smith:1995aa,Jones:1989ab}, whilst a P.A. of polarizations of 123\degr~$\pm$ 4\degr~and 126\degr~$\pm$ 4\degr~at 12 \um~and 14 \um~were measured by \citet{Siebenmorgen:2001ab}. The similar P.A. of polarization from UV to MIR strongly indicates that a single mechanism is responsible for the polarized flux. However, several interpretations of the dominant polarization mechanism have been suggested in the literature: 1) core dominated polarization arising from aligned dust grains in the torus \citep{Siebenmorgen:2001ab}, and 2) dust scattering from the unresolved core \citep[i.e.][]{Smith:1995aa,Gallagher:2005aa}. To investigate the physical mechanism responsible for the IR polarization in the core of Mrk 231, further sub-arcsecond angular resolution IR polarimetric observations are necessary.

In this paper, we aim to disentangle the physical structure that dominates the IR polarization in the unresolved core of Mrk 231. We performed high-angular (0.17 $-$ 0.35 arcsec) resolution imaging polarimetric observations using the 3.1 \um~filter on MMT-Pol in conjunction with the adaptive optics (AO) secondary system on the 6.5-m MMT, and using the 8.7 \um, 10.3 \um, and 11.3 \um~filters on CanariCam on the 10.4-m Gran Telescopio CANARIAS (GTC). We describe the observations and data reduction in  Section 2, and the results are presented in Section 3. A polarization model to account for the IR polarization of Mrk 231 is presented in Section 4. Section 5 presents the discussion of our results, and Section 6 presents the conclusions.



\vspace{-0.5cm}

\section{Observations and Data Reduction}
\label{OBS}

\subsection{MMT-Pol: 3.1 \um~imaging polarimetry}
\label{OBS_NIR}

Mrk 231 was observed on 2015 March 07 using MMT-Pol \citep{Packham:2008aa,Packham:2010aa,Packham:2012aa} in conjunction with the secondary mirror AO system and the f/15 camera on the 6.5-m MMT, Arizona. A dichroic at 15\degr~to the normal before the aperture window of MMT-Pol reflects optical light up to a CCD-based wavefront sensor and passes the IR beam into MMT-Pol at the Cassegrain focus. Thus, Mrk 231 was used for the AO correction at optical wavelengths. MMT-Pol uses a 1024 $\times$ 1024 pixels HgCdTe AR-coated Virgo array, with a pixel scale of 0.043 arcsec pixel$^{-1}$, corresponding to a field of view (FOV) of 44 arcsec $\times$ 44 arcsec. MMT-Pol uses a rectangular focal plane aperture, a half-wave retarder (half wave plate, HWP), and one of two Wollaston prisms. The rectangular focal plane aperture provides two non-overlapping rectangular images with an individual FOV of 20 arcsec $\times$ 40 arcsec. In standard polarimetric observations, the HWP is rotated to four P.A. in the following sequence: 0\degr, 45\degr, 22.5\degr~and 67.5\degr. A Calcite Wollaston is used in the wavelength range of 1$-$2 \um~and a Rutile Wollaston in the 2$-$5 \um~wavelength range.

Observations at the 3.1 \um~($\lambda_{\mbox{\tiny c}}$ = 3.1 \um, $\Delta \lambda =$ 0.1 \um, 50 per cent cut-on/off) narrow band filter of MMT-Pol were performed. This filter provides the best sensitivity within the 3$-$5 \um~instrumental filter set of MMT-Pol, and a new polarimetric measurement from previously published \citep{Jones:1989ab} observations within the 1$-$3 \um~wavelength range. The images were acquired in an ABBA dither pattern with an offset of 10 arcsec in declination, where images per HWP P.A. were taken in each dither position. The position of the vertical axis of the array with the North on the sky was 151\degr~E of N. Frame exposures of 5s per HWP P.A. at each dither position were taken. The observations were performed during windy conditions, where Mrk 231 observations were performed in the direction of the on-coming wind and with a low AO camera frame rate. These conditions degrade the quality of the AO system. Although subsamples of the observations were affected by the weather conditions, only those observations were used with (1) photometric conditions, (2) locked AO in a complete ABBA, and (3) good image quality, providing a total exposure time of 480s. A summary of the observations is shown in Table \ref{table1}.


\begin{table}
\caption{Summary of observations.}
\label{table1}
\begin{tabular}{ccccccc}
\hline
Instrument	&	Date			&	$\lambda_{\mbox{\tiny c}}$	&	On-Source$^{a}$	&	FWHM	\\
			&	(yyyymmdd)	&	(\um)			&	(s)					&				(arcsec)	\\				
\hline	
MMT-Pol		&	20150307		&	3.1			&	480			&	0.14		\\
CanariCam	&	20150302		&	8.7			&	874			&	0.30				\\
				&	20150302		&	10.3			&	728			&	--				\\
				&	20150302		&	11.6			&	1014			&	0.35				\\	
\hline
\end{tabular}

$^{a}$For the MIR observations, the total on-source time was estimated accounting for the positive and negative images, produced by the chop-nod technique, on the array.
\end{table}

The data were reduced using custom {\small IDL} routines, and further analysis using {\small PYTHON}. The difference for each correlated double sample (CDS) pair was calculated, and then sky-subtracted using the closest dither position in time to create a single image per HWP P.A. For each dither position, the ordinary (o-ray) and extraordinary (e-ray) rays, produced by the Wollaston prism, were extracted, and then the Stokes parameters, $I$, $Q$, and $U$ were estimated according to the ratio method \citep[e.g.][]{Tinbergen:2005aa}. The Stokes $I$ were registered and shifted to a common position, where the Stokes $Q$ and $U$ used the same offsets, and then co-averaged to obtain the final $I$, $Q$, and $U$ images. Finally, the degree, P $=\sqrt{Q^2+U^2}$, and the P.A., P.A. $=0.5\arctan{(U/Q)}$, of polarization were computed. During this process, individual photometric and polarimetric measurements were examined at each dither position for high or variable background that could indicate the presence of clouds or electronic problems. Fortunately, no data needed to be removed for these reasons, although some data were removed when the AO guide was unlocked. In the 3.1 \um~filter, we correct by a polarization efficiency of 94 per cent \citep{Shenoy:2015aa}. The instrumental polarization was measured to be 0.09 $\pm$ 0.05 per cent using the unpolarized standard star, SAO133737, during the observing run. Its FWHM was measured to be 0.15 arcsec $\times$ 0.13 arcsec at P.A. = 70\degr, using a Gaussian profile. The measurements of the degree of polarization were corrected for polarization bias using the approach by \citet{Wardle:1974aa}.  

The polarized standard star HD38563A was observed in the 3.1 \um~filter to allow us to estimate the zero-angle calibration of the observations. Frame exposures of 3s per HWP P.A. were taken, with a total of 1 ABBA dither pattern providing a total exposure time of 40s. The zero-angle calibration, $\Delta\theta$, was estimated as the difference of the measured P.A. of polarization of our observations, $\theta_{\mbox{\tiny 3.1\um}}$ = 164\degr~$\pm$ 5\degr, and the P.A. of polarization, $\theta^{\mbox{\tiny W}}_{\mbox{\tiny 3.1\um}} =$ 97\degr $\pm$ 7\degr, provided by \citet{Whittet:1992aa}. The $\theta^{\mbox{\tiny W}}_{\mbox{\tiny 3.1\um}}$ was estimated as the average of the P.A. of polarization from the $B$ to $K$ filters, assuming that the P.A. of polarization of the star does not change with wavelength. Thus, a zero-angle calibration of $\Delta\theta_{\mbox{\tiny 3.1\um}} =$ $\theta^{\mbox{\tiny W}}_{\mbox{\tiny 3.1\um}} - \theta_{\mbox{\tiny 3.1\um}}$ = $-$67\degr~$\pm$ 9\degr~was estimated.

Flux calibration was performed using the polarized standard star SAO133737. We used the magnitudes, 7.700, 7.836, and 7.828 in the $J$ (1.2 \um), $H$ (1.6 \um) and $K$ (2.2 \um) filters by \cite{Cutri:2003aa}. Then, the spectral template of a B1V type star was used for interpolation, with an estimated total flux density of 290 mJy at 3.1 \um. Measured counts from the observations were equated to the flux density of the polarized star. Finally, the factor mJy counts$^{-1}$ was estimated and used in the measurement of the flux density shown in Section \ref{RES}.


\vspace{-0.5cm}

\subsection{CanariCam: mid-infrared imaging polarimetry}
\label{OBS_MIR}

Mrk 231 was observed on 2015 March 02 using the imaging polarimetric mode \citep{Packham:2005aa} of CanariCam \citep{Telesco:2003aa} on the 10.4-m GTC, Spain. Observations were taken within the ESO/GTC large program, and as required by this approved program, all
the data were observed in queue mode under photometric conditions and image quality better than 0.6 arcsec measured at MIR wavelengths from the full width half maximum (FWHM) of the standard stars. CanariCam uses a 320 $\times$ 240 pixel Si:As Raytheon array, with a pixel scale of 0.0798 $\pm$ 0.0002 arcsec pixel$^{-1}$. The imaging polarimetric mode of CanariCam uses a HWP, a field mask and a Wollaston prism. The Wollaston prism and HWP are made with sulphur-free CdSe. The HWP is chromatic, resulting in a variable polarimetric efficiency across the 7.5$-$13 \um~wavelength range, that it has been well-determined \citep{Packham:2008ab}, and the polarimetric mode is usable across this wavelength range. In standard polarimetric observations, the HWP is set in four P.A. in the following sequence: 0\degr, 45\degr, 22.5\degr~and 67.5\degr. The field mask consisted of a series of slots of 320 pixels $\times$ 25 pixels each, corresponding to a FOV of 25.6 arcsec $\times$ 2.0 arcsec, where a total of three slots can be used, providing a non-contiguous total FOV of 25.6 arcsec $\times$ 6.0 arcsec.

The Si2 ($\lambda_{\mbox{\tiny c}}$ = 8.7 \um, $\Delta \lambda =$ 1.1 \um, 50 per cent cut-on/off), Si4 ($\lambda_{\mbox{\tiny c}}$ = 10.3 \um, $\Delta \lambda =$ 0.9 \um, 50 per cent cut-on/off) and Si5 ($\lambda_{\mbox{\tiny c}}$ = 11.6 \um, $\Delta \lambda =$ 0.9 \um, 50 per cent cut-on/off) filters provide the best combination of sensitivity, spatial resolution, and spread of wavelength coverage for the filter set of CanariCam in the 10 \um~atmospheric window. Thus, these filters were used. Observations were made using a standard chop-nod technique to remove time-variable sky background and telescope thermal emission, and to reduce the effect of 1/{\it f} noise from the array/electronics. In all observations, the chop-throw was 8 arcsec, the chop-angle was 90\degr~E of N, and the chop-frequency was 1.93 Hz. The angle of the short axis of the array with respect to the North on the sky (i.e. instrumental position angle, IPA) was 0\degr~E of N, and the telescope was nodded every 46s along the chopping direction. Only one slot with a FOV of 25.6 arcsec $\times$ 2.0 arcsec was used in the observations, with the negative images (produced by the chop-nod technique) within the same slot. For each filter, two observational sets were observed. To improve the signal-to-noise ratio (S/N) of the observations, the negative images on the array were used, providing the total useful on-source time shown in Table \ref{table1}.

Data were reduced using custom \textsc{\small IDL} routines, and further analysis was performed using \textsc{\small PYTHON}. The difference for each chopped pair was calculated and the nod frames then differenced and combined to create a single image per HWP P.A. During this process, all nods were examined for high and/or variable background that could indicate the presence of clouds or variable precipitable water vapour. Fortunately, no data needed to be removed for these reasons. For each observational set, the o- and e-rays, produced by a Wollaston prism were extracted, and then Stokes parameters $I$, $Q$ and $U$ were estimated according to the ratio method. The Stokes $I$ were registered and shifted to a common position, where the Stokes $Q$ and $U$ used the same offsets, and then co-averaged to obtain the final $I$, $Q$, and $U$ images. Finally, the degree and P.A. of polarization were estimated. The instrumental polarization, polarization efficiency and polarization bias were corrected as described by \citet{Lopez-Rodriguez:2014aa,Lopez-Rodriguez:2016ab}.

The polarized young stellar object, AFGL 989, was observed at 8.7 \um~immediately before Mrk 231, with an on-source time of 73s. AFGL 989 allowed us to characterize the polarization observations because it is bright, 99.8 Jy, and polarized, $\sim$2 per cent at 10 \um. The degree of polarization of AFGL 989, corrected by instrumental polarization and polarization efficiency at 8.7 \um, was estimated to be 1.5 $\pm$ 0.2 per cent. Our measurement is in excellent agreement with the degree of polarization of 1.5 $\pm$ 0.7 per cent at 8.7 \um~measured  by \citet[][see fig. 2]{Smith:2000aa}. The zero-angle of P.A. of polarization was calculated as the difference of the P.A. of polarization from our measurement, $\theta =$ 86\degr $\pm$ 6\degr, and \citet{Smith:2000aa}, $\theta_{\mbox{\tiny s}} =$ 105\degr $\pm$ 15\degr, i.e. $\Delta\theta_{\mbox{\tiny 8.7\um}}$ = 19\degr $\pm$ 16\degr~at 8.7 \um. At other wavelengths, the zero-angle was estimated using the wavelength dependence of the P.A. of polarization by \citet[][see fig. 2]{Smith:2000aa}.

Dedicated flux standard stars were not observed. The flux calibration was performed using the N-band spectra observed with CanariCam on the 10.4-m GTC by \citet{Alonso-Herrero:2016aa} with typical uncertainties of 10-15\%. Specifically, flux calibration was performed using the spectral points at 8.7 \um, 10.3 \um, and 11.6 \um~of the 0.4 arcsec wide slit centred at the peak of Mrk 231. The extracted fluxes were 1079 mJy,  936 mJy, and 1465 mJy, respectively. Then, measured counts in a 0.4 arcsec $\times$ 0.4 arcsec simulated slit aperture from our images at each wavelength were equated to the flux densities from the MIR spectra by \citet{Alonso-Herrero:2016aa}. Finally, the factor Jy counts$^{-1}$ was estimated and used in the measurements of the flux densities.


\section{Results}
\label{RES}


\begin{figure*}
\includegraphics[angle=0,trim=0cm 0.5cm 0cm 0cm,scale=0.16]{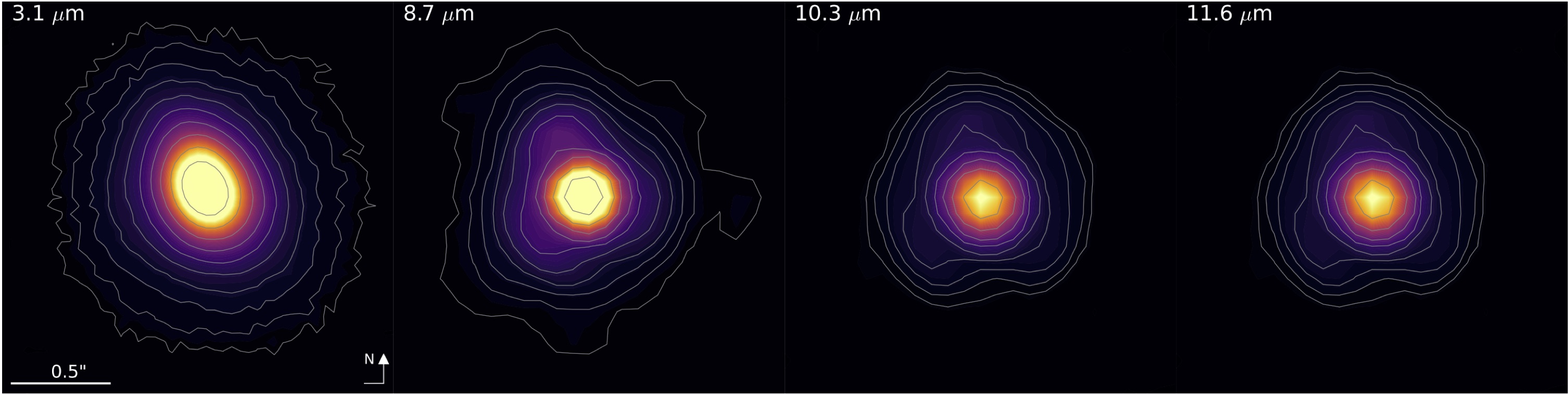}
\caption{From left to right: Total flux images of the central 2 arcsec $\times$ 2 arcsec (1.6 kpc $\times$ 1.6 kpc) in the 3.1 \um~filter of MMT-Pol, and in the 8.7 \um, 10.3 \um, 11.6 \um~filters of CanariCam. In all cases, contours start at 8$\sigma$ and increase as 2$^{n}$, where $n =$ 4, 5, 6, .... The pixel-scale is 0.043 arcsec pixel$^{-1}$ and 0.0798 arcsec pixel$^{-1}$ for MMT-Pol and CanariCam, respectively. At 3.1 \um, the elongation, FWHM = 0.38 arcsec $\times$ 0.29 arcsec at P.A. = 27\degr, does not represent any physical structures of Mrk 231, but this is due to wind buffeting of the telescope and low frame rate of the AO observations. In the MIR, Mrk 231 appears as a point-source with the first Airy ring (0.28 arcsec, 0.33 arcsec, 0.37 arcsec at 8.7 \um, 10.3 \um~and 11.6 \um, respectively) observed at all wavelengths. North is up and east to the left.}
\label{fig1}
\end{figure*}


Figure \ref{fig1} shows the total flux density images of Mrk 231 in the 3.1 \um, 8.7 \um, 10.3 \um~and 11.6 \um~filters. At 3.1 \um, the elongation, FWHM = 0.38 arcsec $\times$ 0.29 arcsec at P.A. = 27\degr, does not represent any physical structures of Mrk 231, but this is due to wind buffeting of the telescope and low frame rate of the AO observations (Section \ref{OBS_NIR}). In the MIR, Mrk 231 appears as a point-source core, showing the first Airy ring (0.28 arcsec, 0.33 arcsec, 0.37 arcsec at 8.7 \um, 10.3 \um~and 11.6 \um, respectively) at all wavelengths. Our observations are in agreement with previous MIR imaging \citep{Matthews:1987aa,Soifer:1987aa,Keto:1992aa}.


\begin{figure*}
\includegraphics[angle=0,trim=0cm 2.5cm 0cm 1.5cm,scale=0.235]{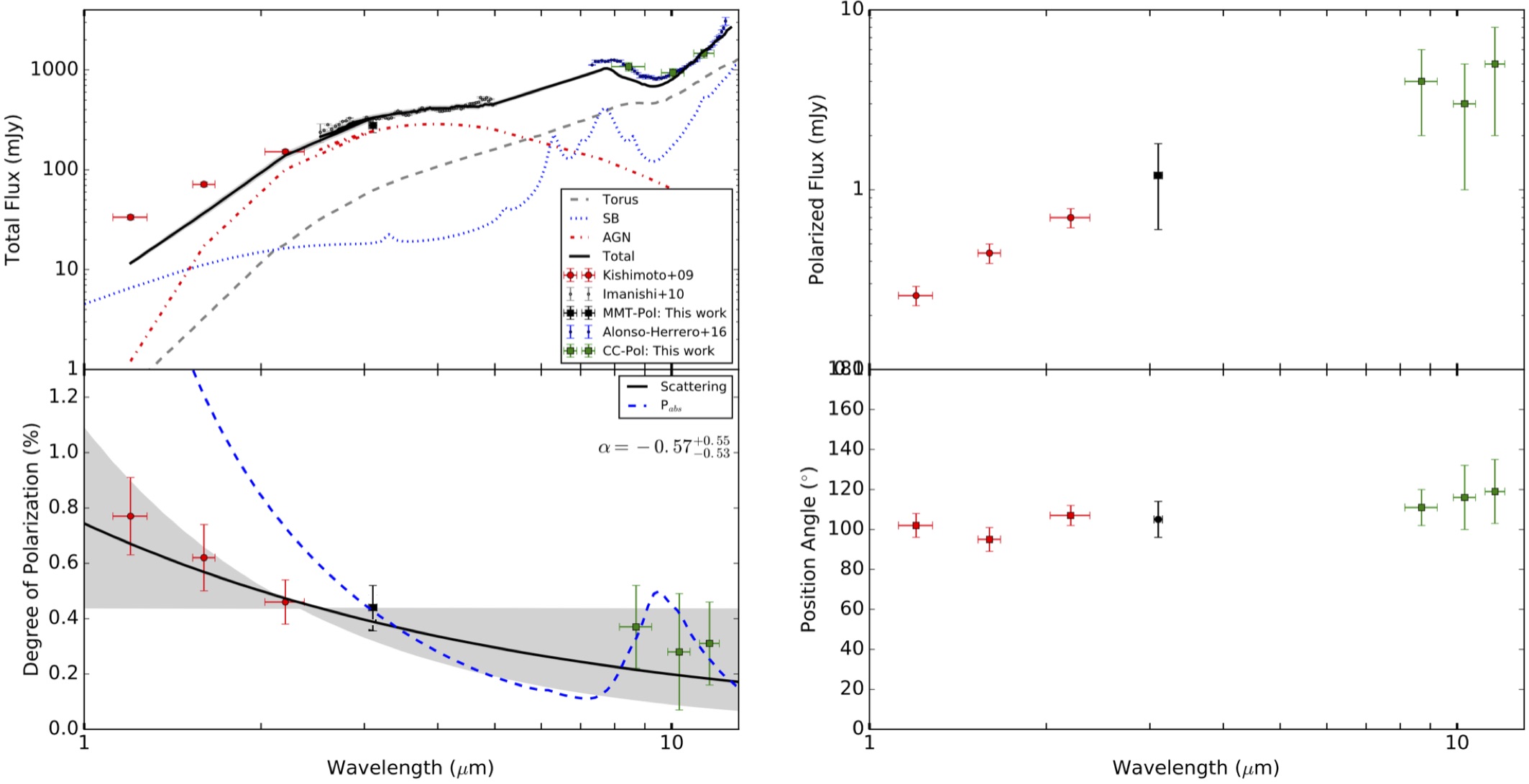}
\caption{Total (top-left)  and polarized (top-right) flux density, degree (bottom-left) and position angle (bottom-right) of polarization of the central 0.5 arcsec (400 pc) region of Mrk 231. The JHK data (red circles) by \citet{Kishimoto:2009aa}, the 3.1 \um~(black square), 8.7 \um, 10.3 \um, and 11.6 \um~data (green squares) as in Table \ref{table2}. The 2$-$2.5 \um~spectrum (grey dots) using AKARI \citep{Imanishi:2010aa}, and 8$-$12 \um~spectrum (blue dots) using GTC/CanariCam \citep{Alonso-Herrero:2016aa} are shown. In total flux: Torus (grey dashed-line), starburst (blue dotted-line), hot dust component (red dotted-dashed line) and total (black solid line) models are shown. In polarization: best power-law, P $\propto$ $\lambda^{\alpha}$, fit with $\alpha =$ 0.57$^{+0.55}_{-0.53}$ and 1$\sigma$ uncertainty (grey shadowed region) is shown. Polarization by dichroic absorption (blue dashed line) normalized to the 3.1 \um~filter, P$_{\mbox{\scriptsize abs}}$, is shown.}
\label{fig2}
\end{figure*}


At all wavelengths, we made measurements of the nuclear photometry and polarimetry in a 0.5 arcsec (400 pc) circular aperture (hereafter, aperture refers to the diameter of a circular aperture) shown in Table \ref{table2}. This aperture optimally measures the nuclear flux density of Mrk 231, and minimizes possible contamination from extended (diffuse) warm nuclear emission and emission from heated dust. In all cases, photometric and polarimetric errors were estimated as the variation of the counts in subsets of the data. Figure \ref{fig2} shows the total flux density, degree of polarization, P.A. of polarization, and polarized flux density in the 1$-$13 \um. The JHK photometry observations (red dots) by \citet{Kishimoto:2009aa}, 2$-$2.5 \um~spectroscopic observations (grey dots) using AKARI \citep{Imanishi:2010aa}, and 8$-$12 \um~spectroscopic observation using GTC/CanariCam by \citet{Alonso-Herrero:2016aa} are shown in Figure \ref{fig2}. To increase the wavelength coverage of the polarimetric observations, allowing us to perform a multi-wavelength polarization analysis (Section \ref{ANA_Model}) in the 1$-$12 \um~wavelength range, Figure \ref{fig2} also includes already published 1$-$3 \um~polarimetric observations from \citet{Jones:1989ab}.

The degree of polarization decreases from 0.77 $\pm$ 0.14 per cent at 1.2 \um~to 0.31 $\pm$ 0.15 per cent at 11.6 \um, although we note that these measurements should be considered as upper-limits. The P.A. of polarization remains fairly constant, $\sim$108\degr, in the 1$-$12 \um, although a tentative increase of the P.A. of polarization of $\sim$15\degr~from 1.2 \um~to 11.6 \um~can be observed. Although the polarization is not formally detected between 8 and 12 \um, there is no evidence for increased polarization near the peak of the silicate feature as might be expected if dichroic absorption by aligned silicate grains made an important contribution at those wavelengths (Section \ref{Pol_Model}). We note that the ISOCAM \citep{Siebenmorgen:2001ab} measurements of the degree of polarization, 8.6 $\pm$ 0.9 per cent and 6.7 $\pm$ 0.9 per cent at 12.0 \um~and 14.3 \um, respectively, are rather different when compared with ours. It is complicated to reconcile the low polarization measured with CanariCam with the ISOCAM results unless any diffuse emission within the ISOCAM beam is very highly polarized. For example, if the 12 \um~polarization is $\sim$8\% in a 10 arcsec beam by ISOCAM, that would require the polarization in the diffuse emission to be very ordered and very high -- probably close to 100\%.  Mrk 231 is a point source and at least 90\% of the flux in a 10 arcsec beam comes from the unresolved core. If 10\% comes from diffuse emission, it would need to be an intrinsic polarization of 90\% to give the claimed ISOCAM results.


\begin{table}
\caption{Photometric and polarimetric measurements of Mrk 231.}
\label{table2}
\begin{tabular}{cccccc}
\hline

	$\lambda_{\mbox{\tiny c}}$		&		Flux density$^{\mbox{\tiny a}}$	&	$P^{\mbox{\tiny b}}$	& 	$P.A.$		&	polarized flux	\\
	(\um)			&		(mJy)				&	(per cent)				&	(\degr)			&	(mJy)	\\
					
\hline
	1.2			&	35$\pm$4				&	0.77$\pm$0.14		&	102$\pm$6		&	0.3$\pm$0.1			\\
	1.6		&	130$\pm$13			&	0.62$\pm$0.12		&	95$\pm$6		&	0.8$\pm$0.3			\\
	2.2		&	192$\pm$40			&	0.46$\pm$0.08		&	107$\pm$5		&	0.9$\pm$0.5			\\	
	3.1				&	278$\pm$42			&	0.44$\pm$0.13		&	105$\pm$9		&	1.2$\pm$0.6							\\
	8.7				&	1079$\pm$108		     &	0.37$\pm$0.15		&	111$\pm$9		&	4$\pm$2	\\
	10.3			     &	936$\pm$94			&	0.28$\pm$0.21		&	116$\pm$16	&	3$\pm$2	\\
	11.6		      	&	1465$\pm$146		     &	0.31$\pm$0.15		&	119$\pm$16	&	5$\pm$3			\\
\hline
\end{tabular} \\
$^{\mbox{\tiny a}}$Photometry: JHK photometry from \citet{Kishimoto:2009aa}; 3.1 \um, 8.7 \um, 10.3 \um, and 11.6 \um~photometry in a 0.5 arcsec from this work. $^{\mbox{\tiny b}}$Polarimetry: JHK polarimetry from \citet{Jones:1989ab}; 3.1 \um, 8.7 \um, 10.3 \um, and 11.6 \um~polarimetry in a 0.5 arcsec aperture from this work.
\end{table}



\vspace{-0.5cm}

\section{Modeling}
\label{ANA_Model}

We aim to account for the observed total nuclear flux and polarization of Mrk231 from 1 \um~to 12 \um. Specifically, we need a polarization mechanism that reproduces the measured degree of polarization decreasing with wavelength and with the fairly constant P.A. of polarization. The total flux and polarization modeling are presented in the following sections.

\vspace{-0.5cm}

\subsection{Total Flux Modeling}

In this section we aim to reproduce the major features of the 1$-$12 \um~SED of Mrk 231 including our observations, JHK observations \citep{Kishimoto:2009aa} taken on UKIRT with a 1 arcsec aperture, 2$-$2.5 \um~spectrum using AKARI \citep{Imanishi:2010aa} with a 7.3 arcsec aperture extraction, and 8$-$12 \um~spectrum using GTC/CanariCam \citep{Alonso-Herrero:2016aa} with a 0.5 arcsec aperture extraction, were used to expand the wavelength coverage. As noted in the introduction, Mrk231 is a point-like source in all wavelength observed to this date, thus the photometric measurement at each wavelength represents the nuclear source at a level of $\ge$90\% independently of the extraction aperture used. As example, high-spatial resolution K and L' band imaging AO observations using Subaru of Mrk 231 show a point like source \citep{Imanishi:2014aa}, indicating that the AKARI observations, within the same wavelength range, describes the photometry of the central source despite the difference in angular resolution. This allows us to construct the nuclear SED of this object. The selection of models have nine free parameters and the particular choices are described in the following subsections.

\vspace{-0.5cm}

\subsubsection{2-phase Clumpy Torus Component}
\label{Torus_model}

It is recognized that clumpy torus models can reproduce the general features from NIR up to sub-mm wavelengths using high-spatial resolution observations \citep[e.g.][]{Ramos-Almeida:2009aa,Ramos-Almeida:2011aa,Alonso-Herrero:2011aa,Lira:2013aa,Ichikawa:2015aa}. A lot of efforts \citep[e.g.][]{Nenkova:2002aa,Nenkova:2008ab,Nenkova:2008aa,Honig:2010aa}, with different levels of sophistication, have been done to obtain a full description of the clumpy torus in AGN. However, these models do not account for 1) both the clumpy dust distribution and an inter-clump dusty component that can co-exist simultaneously at the same scales around the AGN \citep{Stalevski:2012aa,Siebenmorgen:2015aa}, nor 2) different dust grain composition rather than the typical ISM dust grains \citep[MNR;][]{Mathis:1977aa}. We therefore adopt the clumpy torus model\footnote{The SED public library of AGN torus models can be found at: \url{http://www.eso.org/~rsiebenm/agn_models/index.html}} by \citet{Siebenmorgen:2015aa}, who use a 2-phase clumpy torus structure surrounding the central engine. These models assume a dusty structure composed of dusty clumps and an inter-clump medium, in both cases assuming fluffy grains \citep{Kruegel:1994aa}. The 2-phase models allow us to quantitatively estimate the ratio of clumpiness and smoothness dusty structure in Mrk 231 through its five free parameters: 1) the inner radius, $r_{\mbox{\scriptsize in}}$, of the torus, 2) the volume filling factor of the clouds, $\eta$, corresponding to the number of clouds within the 3D model space, 3) the optical depth of the clouds, $\tau_{\mbox{\scriptsize V,cl}}$, 4) the optical depth of the disk midplane, $\tau_{\mbox{\scriptsize V,m}}$, and 5) the viewing angle, $i$, measured from the torus axis.

\subsubsection{Starburst Component}

\citet{Imanishi:2010aa} found 3.3 \um~polycyclic aromatic hydrocarbon (PAH) emission feature using AKARI observations. This result shows evidence of a starburst component within the unresolved core of Mrk 231. To account for a starburst component in the total flux SED, star formation is represented using the \citet{Siebenmorgen:2007aa} models\footnote{The SED model library of starbursts and ULIRGs can be found at: \url{http://www.eso.org/~rsiebenm/sb_models/}} assuming spherical symmetry and an ISM with dust grain properties similar to the Milky Way. Specifically, we consider a starburst radius of 0.35 kpc based on the spatial resolution of our observations, 0.5 arcsec (400 pc), and three free parameters: 1) total luminosity of the starburst, L$_{\scriptsize SB}$, 2) visual extinction, A$_{\mbox{\scriptsize V}}$, and 3) the ratio of the luminosity that is due to OB strs, $\eta_{\mbox{\scriptsize OB}}$, keeping the hot spot density constant at $\eta_{\mbox{\scriptsize HS}} = 10^{4}$ cm$^{-3}$. 

\subsubsection{Hot Dust Component}

We found that the combination of 2-phase clumpy torus model and starburst component do not reproduce the 1$-$12 \um~SED of Mrk 231. Specifically the observations in the 1$-$5 \um~wavelength region are underpredicted. This infrared excess is commonly observed in other Type 1 AGN \citep[e.g.][]{Netzer:2007aa,Mor:2009aa,Mor:2012aa,Stalevski:2012aa,Gallagher:2015aa}. We introduce a dust emission component as a blackbody function with a characteristic temperature of T$_{\mbox{\scriptsize BB}}$. The final characteristic temperature of the blackbody component was estimated as the best fit of a blackbody component in steps of 1 K within the range of 500-1500 K. 

\subsubsection{Total Model}

We obtain the best fit to the observed SED by obtaining the parameter that minimizes the likelihood, $L$, as:

\begin{equation}
L = \prod_{\lambda} \exp\left\{- \frac{(f^{\mbox{\scriptsize obs}}_{\lambda} - f^{\mbox{\scriptsize mod}}_{\lambda} )^2}{2\sigma^{\mbox{\scriptsize obs 2}}_{\lambda}}\right\}
\end{equation}
\noindent
where $f^{\mbox{\scriptsize obs}}_{\lambda}$ is the observed SED, $f^{\mbox{\scriptsize mod}}_{\lambda}$ is the sum of the 2-phase clumpy torus models, starburst models, and the hot dust component, and $\sigma_{\mbox{\scriptsize obs},\lambda}$ is the uncertainties of the observed SED. The models were normalized to the observed SED at 5 \um. The product runs over each observed wavelength, $\lambda$. By simultaneously fitting the 2-phase clumpy torus models, starburst models, and the hot dust component, we obtain the best fit shown in Fig. \ref{fig2}, with the outputs of the free parameters shown in Table \ref{table3}.


\begin{table}
\caption{Model parameters obtained from SED modeling.}
\label{table3}
\begin{tabular}{ccc}
\hline
Component			&	Parameter				&	Value		\\
\hline
Clumpy Torus: 	&	$r_{\mbox{\scriptsize in}}$			& 7.9$\pm$4.2 $\times$ 10$^{17}$ cm	\\
						&	$\eta$											& 3$\pm$2									\\
						&	$\tau_{\mbox{\scriptsize V,cl}}$	& 166$\pm$130								\\
						&	$\tau_{\mbox{\scriptsize V,m}}$	& 369$\pm$287								\\
						&	$i$												& 48$\pm$23\degr							\\
Starburst: 		&	L$_{\mbox{\scriptsize SB}}$ 		& 10$^{12.7\pm0.1}$ L$_{\odot}$ 	\\
						&	A$_{\mbox{\scriptsize V}}$ 			& 36$\pm$5 mag						\\
						&	$\eta_{\mbox{\scriptsize OB}}$	& 40$\pm$10\%							\\
AGN: 				&	T$_{\mbox{\scriptsize BB}}$ 		& 731$\pm$4 K 							\\
\hline
\end{tabular}
\end{table}


We also tried a non-thermal component (i.e. synchrotron) instead of the torus model in the total flux SED. We assumed a power-law as a function of wavelength, but any combination of hot dust component, starburst and non-thermal component cannot reproduce the total flux SED. We can rule out a significant non-thermal component in the unresolved core of Mrk 231. 

Given the number of free parameters using in the SED modeling, we tried empirical starburst templates from the \emph{Spitzer} Wide-area InfraRed Extragalactic survey (SWIRE) template library\footnote{The SWIRE templatea library can be found at: \url{http://www.iasf-milano.inaf.it/~polletta/templates/swire_templates.html}} \citep{Polletta:2007aa} to minimize them and to study the model dependence. This library includes the starburst templates of M82, Arp220, IRAS 20551-4250 and IRAS 22491-1808. For all templates, we found  similar contributions of 2-phase clumpy torus, starburst and hot dust components to those shown in Fig. \ref{fig2}. Moreover, the 1-5 \um~wavelength range was needed to be reproduced by the addition of the hot dust component. We preferred to use the starburst templates by \citet{Siebenmorgen:2007aa} because this library is focused on ultra-luminous galaxies (ULIRGs) and allows us to estimate basic physical parameters of this component.

\subsection{Polarization Modeling}
\label{Pol_Model}

The nearly constant P.A. of polarization from 1 \um~to 12 \um~strongly indicates that a single mechanism is responsible for polarizing the flux. Only one mechanism of polarization within the studied wavelength range is assumed. 

As noted in the introduction, scattering can reproduce the degree and P.A. of polarization from UV to NIR wavelengths of Mrk 231. To test this mechanism, we assume a wavelength dependence in the degree of polarization of $P_{\mbox{\tiny sca}} \propto \lambda^{\tiny {\alpha}}$, with $\alpha$ = [-4, 0]. We use a Bayesian approach to infer the family of solutions that can describe the behavior of the degree of polarization with wavelength. In this case, we use the PyMC3\footnote{PyMC3 is available at: \url{https://github.com/pymc-devs/pymc3}} framework for {\small Python}, which has been successfully applied to a variety of astrophysical problems \citep{Genet:2010aa,Barentsen:2013aa,Wilkins:2013aa,Waldmann:2014aa,Lopez-Rodriguez:2016aa}. PyMC3 is a {\small Python} module for Bayesian statistical modeling and model inference using Markov chain Monte Carlo (MCMC) algorithms. This {\small Python} module uses the No-U-Turn Sampler \citep[NUTS;][]{Hoffman:2011aa}, a Hamiltonian MCMC that avoids the random walk behavior and sensitivity of other MCMC algorithms by taking series of steps through a first-order gradient information. The implementation of this algorithm allows a high-dimensional target distribution to converge more quickly than other methods, such as Metropolis-Hastings or Gibbs sampling. Thus, for the example in this section, the joint distribution using the default settings of PyMC3 was used.  

Using this Bayesian approach, the sampling was carried out using one free parameter, $\alpha$, with an uniform distribution in the range of [-4, 0] as a prior distribution with 25000 samples (with a burn-in length of 1500 samples). The most probable inference is shown in Fig. \ref{fig2}, where the 95\% confidence interval around the median of the family of solutions is shown as grey shadowed regions. The posterior distribution of $\alpha$ with the mode and $\pm 1\sigma$ are estimated as $\alpha = -0.57^{+0.55}_{-0.53}$.

We also investigate the expected IR polarization in terms of dichroic absorption. We follow a similar procedure as described by \citet[][Section 4.3]{Lopez-Rodriguez:2016ab}. In general, a Serkowski curve \citep{Serkowski:1975aa} up to 8 \um~followed by a typical extinction curve of R$_{\mbox{\scriptsize v}} =$ 5.5 in the 8$-$12 \um~is assumed. This composition takes into account the silicate feature at MIR wavelengths, that, if not used, would lead to the expected MIR polarization being underestimated by a factor >10.  This composite absorptive component, P$_{\mbox{\scriptsize abs}}$, is then normalized to the 3.1 \um~polarization measurement (Fig. \ref{fig2}). Although the absorptive component can explain the 3$-$12 \um~polarization (within the uncertainties), it over-predicts the polarization at wavelengths shorter than 3 \um. This behavior was also found by \citet{Jones:1989ab}. If the absorptive component dominates in the core of Mrk 231, it should produce a change in the P.A. of polarization, however this is not consistent with the observed constant P.A. of polarization from UV to MIR, unless the P.A. of the absorptive and scattering components are in the same direction. Thus, we can rule out dichroic absorption component in the core of Mrk 231.


\vspace{-0.5cm}

\section{Discussion}
\label{DIS}

\subsection{Physical components in the central 400 pc}

In the 1$-$5 \um, the dominant emission arises from a hot, 731 $\pm$ 4 K, dust component. 
The relation of dust grain temperature and distance from the central source is r$_{\mbox{\scriptsize pc}} =$ 1.3L$_{\mbox{\scriptsize UV,}46}^{0.5}$ T$_{\mbox{\scriptsize d,}1500}^{-2.8}$ \citep{Barvainis:1987aa} in pc, where L$_{\mbox{\scriptsize UV,}46}$ is the luminosity of the central engine at UV wavelengths in units of 10$^{46}$ erg s$^{-1}$,  T$_{\mbox{\scriptsize d,}1500}$ is the dust grain temperature in units of 1500 K. Using the UV luminosity of L$_{\mbox{\scriptsize UV}}$ = (2.74 $\pm$ 0.4) $\times$ 10$^{44}$ erg s$^{-1}$ \citep{Munoz-Marin:2007aa}, a distance of r$_{\mbox{\scriptsize pc}} =$ 1.6 $\pm$ 0.1 pc is estimated. The mass of dust associated with the hot dust component can be estimated approximately, as it depends on the unknown dust grain composition and size distribution. The luminosity of individual graphite grains is given by  L$_{\nu,\mbox{\scriptsize IR}}^{\mbox{\scriptsize d}} =$ 4$\pi a^{2} \pi Q_{\nu} B_{\nu}$($T_{\mbox{\scriptsize d}}$) erg s$^{-1}$ Hz$^{-1}$ \citep{Barvainis:1987aa}, where $a$ is the dust grain radius, $Q_{\nu} = q_{\mbox{\scriptsize IR}} \nu^{\gamma}$ is the absorption efficiency of dust grains, and $B_{\nu}$($T_{\mbox{\scriptsize d}}$) is the Planck function for a dust grain of temperature $T_{\mbox{\scriptsize d}}$. We take $a =$ 0.05 \um, and in the NIR, $q_{\mbox{\scriptsize IR}} =$ 1.4 $\times$ 10$^{-24}$ and $\gamma =$ 1.6, yield  $Q_{\nu} =$ 0.033 at 3.1 \um. Using the observed flux density at 3.1 \um~of 278 $\pm$ 42 mJy and 260 mJy extracted from the hot dust component, we estimated N$_{\mbox{\scriptsize gr}} =$ 1.12 $\times$ 10$^{49}$ number of grains in the 0.5 arcsec (400 pc) core of Mrk 231. Given typical interstellar grains \citep[MNR,][]{Mathis:1977aa} a grain density of $\rho =$ 2.26 g cm$^{-3}$ \citep{Barvainis:1987aa} is assumed, and the mass of dust in the 0.5 arcsec (400 pc) aperture is estimated to be M$_{\mbox{\scriptsize d}} =$ N$_{\mbox{\scriptsize gr}}$ V$_{\mbox{\scriptsize d}}$ $\rho$ = 6.7 $\pm$ 2.2 M$_{\odot}$. 

As the torus emission, which includes its own hot dust emission from directly radiated clouds, is already accounted for the 2-phase torus component, and the BLR is dust-free within typical scales of 0.3 pc, we attribute the hot dust component at a distance of r$_{\mbox{\scriptsize pc}} =$ 1.6 $\pm$ 0.1 pc to the pc-scale polar regions. This hot dust component can also be part of the known outflows \citep{Feruglio:2015ab,Morganti:2016aa,Veilleux:2016aa} in Mrk231. This hot dust component has been previously suggested 1) to reproduce the NIR emission in a sample of type 1 AGN \citep{Netzer:2007aa,Mor:2009aa,Mor:2012aa}, and 2) as a dusty wind launched from the inner part of the torus that account for most of the IR emission in several AGN \citep{Honig:2010aa,Honig:2012aa,Honig:2013aa,Gallagher:2015aa}.

A starburst component within the 0.5 arcsec (400 pc) is required to reproduce the features observed in the 2$-$5 \um~AKARI observations \citep{Imanishi:2010aa}. In the 8$-$12 \um, the moderate silicate feature in absorption is reproduced by the starburst component. Although the torus model alone cannot reproduce the 9.7 \um~silicate feature, a contribution of the torus is necessary to reproduce the SED. Our findings are in agreement with a recent study by \citet{Alonso-Herrero:2016ab}, they found that the central 0.4 arcsec MIR emission of Mrk 231 can be explained as a contribution of 90\% of AGN emission, 6\% of star formation and 4\% of stellar contribution from the host galaxy. \citet{Alonso-Herrero:2016ab} performed a spectral decomposition using \textsc{DeblendIRS} \citep{Hernan-Caballero:2015aa} with the same spectrum used in our study. Based on the 2-phase torus model, the torus has an inclination of 48 $\pm$ 23\degr, consistent with a Type 1 AGN, with a filling factor of $\eta$ = 3 $\pm$ 2 clouds within the full 3D model space. This indicates that the 2-phase clumpy torus is dominated by a disk-like dusty component (smooth torus). 

We interpret the several physical components of Mrk 231 within the 0.5 arcsec (400 pc) as a combination of thermal components in the following manner: 1) a hot dust component directly irradiated by the central engine and emitting most of the emission in the 1$-$5 \um~wavelength range, which it is located in the pc-scale polar regions, 2) an optically thick, smooth and dusty structure (torus) with an inclination of 48 $\pm$ 23\degr~surrounding the central engine, and 3) a starburst component, with a total luminosity of 10$^{12.7\pm0.1}$ L$_{\odot}$, at a distance $<$400 pc, is responsible of the 3.3 \um~ PAH emission and the 9.7 \um~absorption feature.

\subsection{Mechanism of polarization}

The most plausible explanation of the measured polarization is that the polarization arises from scattering off dust grains in the unresolved core of Mrk 231. We found a wavelength dependence of the polarization as $P \propto \lambda^{\alpha}$, with $\alpha = -0.57^{+0.55}_{-0.53}$. This component can explain the increase of the degree of polarization as high as $\sim$20\% in the UV wavelength range \citep{Smith:1995aa,Gallagher:2005aa}. Given that the polarization is observed from UV to IR, the scattered material is composed by small, $\le$0.1 \um, dust grains.

Based on our total flux SED modeling, the hot, 731 $\pm$ 4 K, dust structure located in the pc-scale polar regions can explain the scattered material producing the polarization of Mrk 231. As Mrk 231 is a Type 1 AGN, only the hot dust component in our LOS is observed, which favors the asymmetric distribution in the polar regions. This interpretation is consistent with previous studies \citep[i.e.][]{Schmidt:1985aa,Smith:1995aa,Gallagher:2005aa} suggesting that the nuclear polarization arises from a dust component close to the central engine of Mrk 231. Recent UV observations by \citet{Veilleux:2016aa} found broad blue-shifted near-UV absorption lines overlapping with far-UV emission lines, which indicates the presence of a dusty high-density and patchy broad absorption line screen covering $\sim$90\% at a distance of $<$2$-$20 pc. \citet{Feruglio:2015ab} found an ultra-fast nuclear wind with dominant components along the south-west to north-east direction, which suggest a wide-angle biconical geometry. It is tentative that these dusty winds, which are almost perpendicular \citep[Fig. 17,][]{Feruglio:2015ab,Morganti:2016aa} to our measured P.A. of polarization, can produce the scattered polarization. Alternatively, a direct view of a sub-parsec warped disk in the central region of Mrk 231 can also give rise to polarization by scattering \citep[e.g. Cygnus A,][]{Tadhunter:2000aa} due to its asymmetric geometry, however further details about this geometry is very complex to provide and presents a multi-degeneracy problem.
 
Could other polarization mechanisms contribute to our observations? Synchrotron emission as the dominant polarization can be immediately eliminated due to 1) the lack of a non-thermal component in the total flux SED, 2) the variation of the polarization as a function of wavelength, and 3) by the similar polarization of the emission lines and surrounding continuum in the optical and UV \citep[e.g.][]{Smith:1995aa}. Interstellar polarization can also be ruled out due to 1) the overall polarization strength expected to be $<$ 0.1\%, and 2) that the polarization increases up to UV wavelengths which is not consistent with the typical Serkowski curve \citep{Serkowski:1975aa} by dichroic absorption.



\section{Conclusions}
\label{CON}
 
We present, for first time, the combination of sub-arcsecond imaging polarimetric observations of Mrk 231 at 3.1 \um~using MMT/MMT-Pol and 8.7 \um, 10.3 \um~and 11.6 \um~using GTC/CanariCam. We found a decrease in the degree of polarization with increasing wavelength, while the P.A. of polarization remains constant. To put physical constraints on the polarization mechanism, we combine our total flux and polarimetric observations with previous 1$-$2 \um~imaging observations, and 2$-$5 \um~spectroscopic observations using AKARI. 

Based on our modeling of the 1$-$12 \um~total flux SED, several physical structures in the central 400 pc of Mrk 231 can be explained as follows: 1) a hot dust component directly irradiated by the central engine and emitting most of the emission in the 1$-$5 \um~wavelength range, which it is located in the pc-scale polar regions, 2) an optically thick, smooth and dusty structure (torus) with an inclination of 48 $\pm$ 23\degr~surrounding the central engine, and 3) a starburst component, with a total luminosity of 10$^{12.7\pm0.1}$ L$_{\odot}$, at a distance $<$400 pc, is responsible of the 3.3 \um~ PAH emission and the 9.7 \um~absorption feature.

The most plausible explanation of the measured polarization in the core of Mrk 231 is that it arises from scattering off hot dust grains from an asymmetric distribution within the unresolved core. The hot, 731 $\pm$ 4 K, dust structure located in the pc-scale polar regions can explain the scattered material producing the polarization of Mrk 231.

\section*{Acknowledgments}

We would like to thank Dr. Chiara Feruglio for their useful comments, which improved the paper signicantly. Based on observations made with the Gran Telescopio CANARIAS (GTC), instaled in the Spanish Observatorio del Roque de los Muchachos of the Instituto de Astrof\'isica de Canarias, in the island of La Palma. Based on observations made with MMT-Pol on the 6.5-m MMT. C.P. acknowledge support from the University of Texas at San Antonio. C.P. acknowledges support from NSF-0904421 grant. C.P. and T.J.J acknowledge support from NSF-0704095 grant. A.A.-H. acknowledges financial support from the Spanish Plan Nacional de Astronom\'ia y Astrofis\'ica under grant AYA2012-31447, which is party funded by the FEDER program, and financial support from the Spanish Ministry of Economy and Competitiveness (MINECO) under the 2011 Severo Ochoa Program MINECO SEV-2011-0187. P.E. from grant AYA2012-31277, and L.C. from grant AYA2012-32295. C.R.A. acknowledges the Ramón y Cajal Program of the Spanish Ministry of Economy and Competitiveness (RYC-2014-15779). N.A.L. is supported by the Gemini Observatory, which is operated by the Association of Universities for Research in Astronomy, Inc., on behalf of the international Gemini partnership of Argentina, Brazil, Canada, Chile, and the United States of America. R.N.  acknowledges support by FONDECYT grant No. 3140436

\bibliographystyle{mnras}
\bibliography{Mrk231_paper}


\bsp	
\label{lastpage}
\end{document}